\documentclass[a4paper,11pt]{article}
\usepackage{amsmath,amssymb,bm}
\usepackage[T1]{fontenc}
\usepackage{mathtools}
\usepackage{dsfont}
\pdfoutput=1
\usepackage[english]{babel}
\usepackage{graphicx} 
\usepackage{booktabs} 
\usepackage{color}
\definecolor{aquamarine}{rgb}{0.5, 1.0, 0.83}
\definecolor{ao(english)}{rgb}{0.0, 0.5, 0.0}
\definecolor{armygreen}{rgb}{0.29, 0.33, 0.13}
\definecolor{awesome}{rgb}{1.0, 0.13, 0.32}
\definecolor{ballblue}{rgb}{0.13, 0.67, 0.8}
\definecolor{bittersweet}{rgb}{1.0, 0.44, 0.37}
\definecolor{blue}{rgb}{0.0, 0.0, 1.0}
\definecolor{brinkpink}{rgb}{0.98, 0.38, 0.5}
\definecolor{ballblue}{rgb}{0.13, 0.67, 0.8}
\definecolor{brightturquoise}{rgb}{0.03, 0.91, 0.87}
\definecolor{blue-green}{rgb}{0.0, 0.87, 0.87}
\definecolor{caribbeangreen}{rgb}{0.0, 0.8, 0.6}
\definecolor{cyan}{rgb}{0.0, 1.0, 1.0}
\definecolor{amber(sae/ece)}{rgb}{1.0, 0.49, 0.0}
\graphicspath{ {images/} }

\author{J\"{u}rg Fr\"{o}hlich}
\title{A Brief Review of the ``$ETH$- Approach to Quantum Mechanics''}

\begin{document}

\maketitle

\begin{abstract}
To begin with, some of the conundrums concerning Quantum Mechanics and its interpretation(s) are recalled. Subsequently, a sketch of the ``ETH-Approach to Quantum Mechanics'' is presented. This approach yields a logically coherent quantum theory of ``events'' featured by isolated physical systems and of direct or projective measurements of physical quantities, without the need to invoke ``observers''. It enables one to determine the stochastic time evolution of states of physical systems. 
We also briefly comment on the quantum theory of indirect or weak measurements, which is much easier to understand and more highly developed than the theory of direct (projective) measurements. 
A relativistic form of the ETH-Approach will be presented in a separate paper. 
\end{abstract}

\tableofcontents

\section{Introduction -- comments on the foundations of Quantum Mechanics, and purpose of paper}\label{Intro}

Let me start with a few general remarks: I consider it to be an intellectual scandal that, nearly one hundred years after the discovery of matrix mechanics by \textit{Heisenberg, Born, Jordan} and \textit{Dirac}, many or most professional physicists -- experimentalists and theorists \mbox{alike --} admit to being confused about the deeper meaning of Quantum Mechanics ($QM$), or are trying to evade taking a clear standpoint by resorting to agnosticism or to overly abstract formulations of $QM$ that often only add to the confusion. Attempts to replace $QM$ by some alternative deterministic theory, one that does not have a ``measurement problem'', yet reproduces important predictions of $QM$, do not appear to have been very successful, so far. Unfortunately, most physicists have prejudices preventing them from taking a fresh, unbiased look at the subject, and discussions of the foundations of $QM$ tend to be surprisingly emotional. \textit{I feel it is time to change this situation!}

My own interests in the foundations of Quantum Mechanics were aroused in courses on $QM$ taught by \textit{Klaus Hepp} and \textit{Markus Fierz} in the late sixties of the past century, which I took as an undergraduate student. I suppose that most serious students of Physics develop such interests during their first courses on $QM$. But I felt that the subject had better remain a hobby until later in my career. Not least because of the appearance of partly contradictory novel \textit{``interpretations of $QM$''}, all of which left me unsatisfied, (see, e.g., \cite{Griffiths, Durr-Teufel}, and \cite{FS-Vienna} for a brief survey), my views of the foundations of $QM$ actually remained quite confused until a little more than ten years ago; (which did not prevent me from giving talks about the subject -- some with modest impact -- in numerous places). But when I was approaching mandatory retirement I felt an urge to clarify my understanding of some of the subjects I had had to teach to my students for thirty years -- thermodynamics, effective dynamics (in particular Brownian motion), and, foremost, the foundations of $QM$; see \cite{Abou-Salem-F, DeR-Fr, Gang-F, B-DeR-F} and references given there, the last two papers having some relevance for the foundations of $QM$.\footnote{I think it is more appropriate to speak of the ``foundations of $QM$'', rather than ``interpretations of $QM$''. We have to understand what $Q M$ tells us about Nature, \textit{what it means} - once this is accomplished, the correct interpretation of the theory will come almost automatically.} At the beginning of 2012, my interests in this subject became more serious, and I pursued them in joint efforts with my last PhD student, \textit{Baptiste Schubnel}. Later, some further colleagues got interested in our efforts, including \textit{M. Ballesteros, Ph. Blanchard, N. Crawford, J. Faupin} and \textit{M. Fraas}, who collaborated with us in changing configurations. At this point, I wish to thank my collaborators for their support in this endeavor, as well as quite a few colleagues -- too many to mention all of them -- who were willing to listen to me and discuss ideas on basic questions concerning the foundations of $QM$ with me. \textit{D. D\"{u}rr} and \textit{S. Goldstein} deserve my thanks for the encouragement and understanding they have provided.

In this paper, I present a sketch of the ``$ETH$-Approach to Quantum Mechanics'' \cite{FS-Prob-Theory, BFS-forks, Schubnel-thesis}. The $ETH$-Approach is supposed to lay the foundations of a logically coherent quantum theory of \textit{``events''} \cite{Haag} and of \textit{direct} or \textit{projective measurements} of physical quantities (serving to record ``events'') that does not require invoking any ``deos ex machina'', such as ``observers''; (see also \cite{Durr-Teufel}). I have given quite a few talks about this new approach. Technical details have been presented in a short course taught at Les Diablerets, in January of 2017 \cite{Les-Diablerets}, and in \cite{FFS, FS-state-prep}. Our work has profited from ideas proposed by the late \textit{Rudolf Haag} \cite{Haag}, from a paper of \textit{D. Buchholz} and the late \textit{J. E. Roberts} \cite{Buchholz}, and from discussions with Buchholz. In completing this paper I enjoyed receiving feedback from a very careful referee who found many typos and pointed out various unclear statements. A form of the $ETH$-Approach compatible with Einstein causality and Relativity Theory is sketched in \cite{Fr}. But a comprehensive review of our work has not been written, yet. 

Wide-spread recent interest in foundational problems surrounding $QM$ has been triggered by problems in quantum information theory and by the 2012 Nobel Prize in Physics awarded to \textit{S. Haroche} \cite{Raimond-Haroche} and \textit{D. Wineland}. Their discoveries, as well as results described in \cite{Bauer-Bernard, Maassen-Kummerer}, and references given there, have influenced some of our own work on the theory of indirect measurements in $QM$, which has appeared in \cite{BFFS, BCFFS-1, BCFFS-2} and is briefly sketched at the end of this paper. The theory of indirect (``non-demolition-'' and ``weak-'') measurements is quite well developed and clear, \textit{assuming} one understands what ``events'' and ``direct measurements and observations'' are, specifically \textit{direct} observations of ``probes'' used to \textit{indirectly} retrieve information on physical systems. The theory of ``events'' and of ``direct (projective) measurements'' actually constitutes the deep and controversial part of the foundations of $QM$, and it is a novel approach to this theory that I intend to outline in this paper. 

\section{Standard formulation of Quantum Mechanics and its shortcomings}
In our courses on Quantum Mechanics, physical systems, $S$, are often described as pairs, $(\mathcal{H}, U)$, of a Hilbert space, $\mathcal{H}$, of pure state vectors and a propagator, $U$, consisting of unitary operators 
$\big(U(t,t')\big)_{t,t' \in \mathbb{R}}$, acting on $\mathcal{H}$ \textit{seemingly} describing the time-evolution of state vectors in $\mathcal{H}$ from time $t'$ to time $t$. The state space $\mathcal{H}$ of physically realistic systems tends to be infinite-dimensional (but separable). Alas, all infinite-dimensional separable Hilbert spaces are isomorphic, and the data invariantly encoded in the pair $(\mathcal{H}, U)$ do not tell us anything interesting about the physics of $S$, beyond spectral properties of the operators $U(t,t')$, (i.e., ``energy levels''); and they lead one to the mistaken impression that $QM$ might be a \textit{linear} and \textit{deterministic} theory -- alas, one that is entirely inadequate to describe events and the outcome of observations and measurements.

We must therefore clarify what should be added to the formalism of $QM$ in order to capture its fundamentally probabilistic nature and to arrive at a mathematical structure that enables one to describe {\bf{physical phenomena}} (``events'') in  \textit{isolated open} systems $S$, without a need to appeal to the intervention of ``observers'' with ``free will'' -- as is done in the conventional \textit{``Copenhagen Interpretation of QM''} -- or to assume that other ``ghosts'' not intrinsic to the theory come to our rescue.

\underline{\bf{Isolated open systems}}: An \textit{isolated system} $S$ is one that, for all practical purposes, does not have any interactions with its complement, i.e., with the rest of the Universe; meaning that, for periods of time much longer than the time of monitoring it, interactions between the degrees of freedom of $S$ and those of its complement can be neglected in the description of the Heisenberg-picture time evolution of operators. This does, however, \textit{not} exclude that the \textit{state} of $S$ may be entangled with the state of its complement. The special role played by isolated systems in discussions of the foundations of $QM$ stems from the fact that, \textit{only for an isolated system, S}, the time evolution in the {\bf{Heisenberg picture}} of arbitrary operators acting on $\mathcal{H}$ is given by conjugation with the unitary propagator, $U$, of $S$, (determined by its \textit{Hamiltonian}).
An isolated system $S$ is called \textit{open} if it can emit modes to the outside world (the complement of $S$) that eventually cannot be recorded, anymore, by any devices belonging to $S$, yet can be in a state entangled with the state of $S$ after emission. The reader may think of photons or gravitons emitted by an isolated system $S$ that escape from detection by any devices in $S$. (See also Definition 1, below.)\hspace{1cm} $\square$
 
Physical quantities characteristic of a system $S$ are described by certain self-adjoint linear operators, $X=X^{*}$, acting on $\mathcal{H}$. This feature is common to \textit{all physical theories} used at present.\footnote{In classical theories, these operators generate an \textit{abelian} ($C^{*}$-) algebra, and time evolution is given by a $^{*}$-automorphism group of this algebra generated by a vector field on its spectrum; while, in $QM$, the algebra generated by operators representing physical quantities (and events) is \textit{non-commutative}, and time evolution is given by a $^{*}$-automorphism group of such an algebra \textit{only} if the system is \textit{isolated}.} The Copenhagen Interpretation of Quantum Mechanics then stipulates that there are \textit{``observers''} with ``free will'' who can decide to measure such physical quantities arbitrarily quickly, at arbitrary times, and at an arbitrary rate. It is argued that the time evolution of physical states of $S$ is determined by its unitary propagator $U$, which solves a (deterministic) \textit{Schr\"{o}dinger equation}, \textit{except} when a measurement of a physical quantity represented by an operator $X=X^{*}$ is made: Immediately after the measurement of $X$ the state of $S$, according to the Copenhagen Interpretation, is in an eigenstate of $X$ corresponding to the measured value of $X$. If this value is not recorded, one is advised to use a density matrix describing an \textit{incoherent} superposition of eigenstates of $X$, chosen in accordance with \textit{Born's Rule}, to describe the future evolution of $S$. 

For a variety of reasons, this is not a satisfactory recipe for how to apply $QM$ to describe physical phenomena! One might want to view the evolution of states in the presence of measurements, as described in the Copenhagen Interpretation of $QM$, as some kind of stochastic process. But the problem is that one is dealing with a stochastic process that does \textit{not} have a classical state space, and that it is \textit{transition amplitudes}, rather than \textit{transition probabilities}, that are given by matrix elements of an operator (the propagator $U$) satisfying a group composition law, i.e., a kind of Chapman-Kolomogorov equation.\footnote{It is advocated by certain groups of people that the problem arising from this fact can be remedied by invoking the phenomenon of ``decoherence'' and appealing to the ``consistency'' of histories of events \cite{Griffiths}. But I find the arguments supporting this point of view unconvincing.} According to the Copenhagen Interpretation, predicting/determining the transition probabilities describing the stochastic time-evolution of states of $S$ in the presence of repeated measurements would apparently require knowing what kind of physical quantities are measured by the intervention of ``observers'', and at what times these measurements are made. For, any intermediate intervention of an ``observer'' destroys ``interference effects''; and hence it seemingly affects the value of the transition probability between an initial state of $S$ in the past and a target state in the future, \textit{even} if a sum over \textit{all possible} outcomes of the intermediate intervention is taken.\footnote{This is the case unless perfect ``decoherence'' holds.} Without complete information on all intermediate measurements performed on $S$, which, in the Copenhagen Interpretation, is \textit{not} provided by the theory, reliable predictions of future states of the system and of future expectation values of physical quantities become impossible. As a result, the Copenhagen Interpretation renders $QM$ nearly ``unpredictive'' -- even though, by experience, it is a heuristic framework supplementing $QM$ that works well for many or most ``practical purposes'', because, much of the time (in particular when using a scattering matrix), one is interested in predicting the outcome of only \textit{a single} measurement. The situation is hardly improved in a definitive way by resorting to concepts such as ``decoherence'' and interpretations such as ``consistent histories'' \cite{Griffiths}, ``many worlds'', etc.. (See \cite{Bell-book, Durr-Teufel} for further information.)

Before proceeding to describe the ``$ETH$-Approach'', I recall an argument, presented in detail in \cite{FFS}, that shows that the Schr\"{o}dinger equation does \textit{not} describe the time evolution of \textit{states} of systems in the presence of ``events'' or ``measurements'', \textit{assuming} that the usual correlations between the outcomes of Bell-type measurements, claimed to be confirmed in many experiments, hold.

We consider the following Gedanken-Experiment \cite{FFS}, which, ultimately, will show that {\textit{time evolution of states in $QM$ is intrinsically stochastic, in spite of the deterministic nature of the Schr\"{o}dinger equation.}}
\begin{center}
\includegraphics[width=7.9cm]{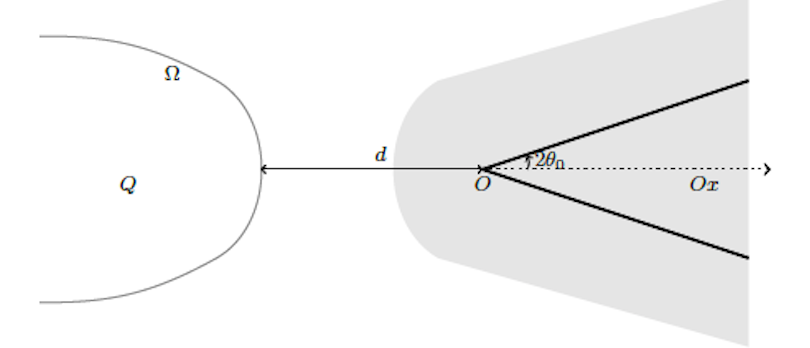}\\
{\small{$\uparrow$}}\hspace{4cm} {\small{$\uparrow$}}\\
$Q= \text{sub-system ``confined'' to } \Omega$\, \hspace{0.3cm} Particle $P$ propagating into shaded cone\\
\textit{Figure 1}
\end{center}
We prepare the system $Q\vee P$ in a state with the property that particle $P$ propagates into the shaded cone opening to the right, except for tiny tails leaking beyond this region, while the degrees of freedom of $Q$ remain confined to a vicinity of the region $\Omega$ in the complement of the shaded cone, except for tiny tails. Thanks to cluster properties, expectation values of the Heisenberg-picture time evolution of physical quantities, such as spin, momentum, etc. referring to $P$ in this state then turn out to be essentially \textit{independent} of the time evolution of the degrees of freedom of $Q$. In other words, interaction terms in the Hamiltonian of 
the system coupling $P$ to $Q$ can be neglected. This is discussed in much detail in \cite{FFS}.

More concretely, we study the following system:
\begin{center}
\includegraphics[width=9.1cm]{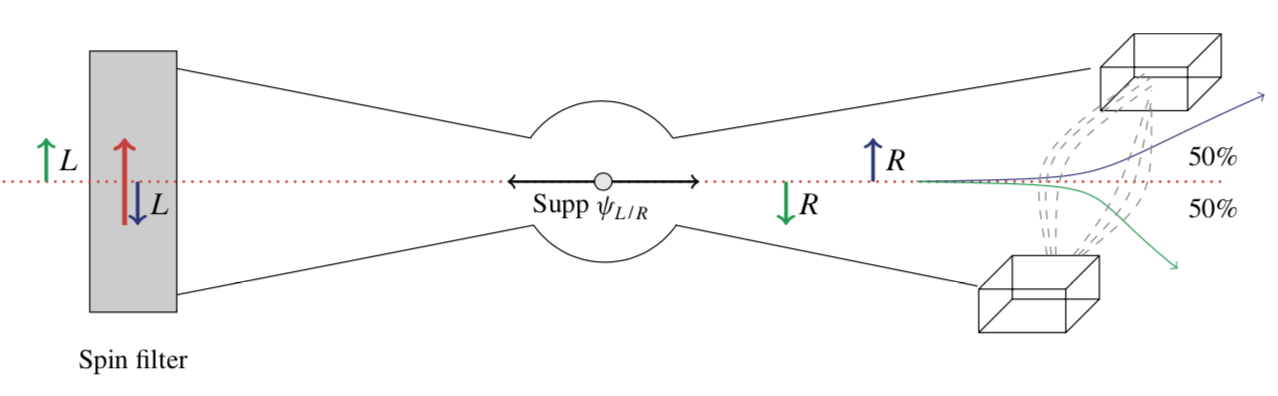}\\
\small{
Q:=$\lbrace$spin filter\, $\vee$\, particle\, P'$\rbrace$ \hspace{0.45cm} cone opening to right:= ess. supp of orbital \mbox{wave function of P\,}\\
}
\vspace{0.1cm}
\textit{Figure 2}
\end{center}

{\bf{Temporary assumptions}} (leading to a contradiction):
\begin{itemize} 
\item{$P$ and $P'$: Two spin-$\frac{1}{2}$ particles prepared in a \underline{spin-singlet initial state}, $\psi_{L/R}$, localized, initially, in the central region shown in Figure 2; the orbital wave function of $P$ is chosen such that $P$ propagates into the cone opening to the right (except for very tiny tails) and that it will eventually undergo a Stern-Gerlach spin measurement, while the orbital wave function of $P'$, an electron, is chosen such that this particle propagates into the cone opening to the left, with only very tiny tails leaking beyond this cone into the half-space to the right of the spin filter. (One may assume, for simplicity, that there are no terms in the total Hamiltonian of the system describing direct interations between $P$ and $P'$.)
The spin filter (e.g., a spontaneously magnetized metallic film) is prepared in a \underline{poorly known} initial state.}
\item{\textit{The dynamics of the} \underline{state} \textit{of the} \underline{total} \textit{system is assumed to be fully determined by a Schr\"{o}dinger equation given by a concrete self-adjoint Hamiltonian containing only short-range interaction terms.} In particular, the initial state of the total system (consisting of the spin filter, the two particles and possibly some Stern-Gerlach equipment serving to measure a component of the spin of particle $P$) is assumed to \underline{determine} whether particle $P'$ will pass through the spin filter, or not, (given that the initial state of $P' \vee P$ is a spin-singlet state, with $P'$ and $P$ moving into \underline{opposite} cones).} Since it is assumed that a Schr\"{o}dinger equation determines the evolution of \textit{states} of this system, the Schr\"{o}dinger picture and the Heisenberg picture are equivalent.
\item{\textit{Correlations between the outcomes of spin measurements of $P'$ and of $P$ are assumed to be those predicted by standard quantum mechanics}, (relying on the ``Copenhagen interpretation'' and apparently confirmed in many experiments): We first note that if $P'$ passes through the spin filter then its spin is ``up'', (i.e., aligned with the majority spin of electrons in the spin filter), if it does not pass through the filter, (i.e., if it hops into a vacant state localised inside the spin filter), its spin is ``down''. The second assumption stated above then says that, whether $P'$ passes through the filter, or not, is determined by the \textit{inital state} of the total system and by solving a \textit{deterministic Schr\"{o}dinger equation}. In addition to the two assumptions already stated, we also assume that \underline{if the spin of $P'$ is measured to be ``up'' the \,spin of $P$ is measured to be ``down}'' (for example, in a Stern-Gerlach experiment involving a magnetic field parallel to the majority spin of the spin filter), and if the spin of $P'$ is ``down'' then the spin of $P$ is ``up''.}
\end{itemize}

Next, we recall the\\
\underline{Fact:}
Expectation values of observables (such as spin, momentum, etc.) referring to particle $P$ in the state of the system described above are \textit{independent} of the degrees of freedom of 
$Q:= \lbrace P'\vee \text{ spin filter}\rbrace$, for arbitrarily long times, up to very tiny corrections. Thus, to a very good approximation, their evolution can be assumed to be given by free-particle dynamics. This is a consequence of our choice of an initial state (propagation properties of the orbital wave functions of $P$ and $P'$) and of cluster properties of the time evolution -- as shown in \cite{FFS}. \\
It follows that, to a very good approximation, the \textit{spin of $P$ is \textit{conserved} before it is measured} \,\,$\Rightarrow$
\begin{center}
\underline{Expectation value of spin of $P \approx 0, \forall$ times before measurement time,}
\underline{\textit{independently} of the evolution of $Q= \lbrace P' \vee \text{  spin filter}\rbrace$!}
\end{center}

But \textit{this {\bf{contradicts}} the third (last) assumption stated above:} The first two assumptions imply that the values of the $z$-component of the spin of $P'$ measured with the help of the spin filter do apparently \textit{not} introduce any bias in the outcomes of measurements of the $z$-component of the spin of $P$. In other words, the second assumption stated above is incompatible with the \textit{Bell-type ``non-locality''} of Quantum Mechanics, as expressed in the third assumption stated above. \\
This argument is robust, in the sense that it suffices to assume that correlations between measurements of a component of the spin of $P'$ and a component of the spin of $P$ are fairly close to those predicted by the Bell-type non-locality described in the third assumption.\\

\underline{Conclusion:} If the third assumption holds true then the quantum-mechanical time evolution of \textit{states} of physical systems in the presence of measurements (or ``events'') is \textit{not} given by a deterministic Schr\"{o}dinger equation, and the equivalence of the Heisenberg picture and the Schr\"{o}dinger picture apparently fails. Quantum Mechanics appears to be intrinsically probabilistic (and ``non-local'', in the sense of Bell-type correlations -- which does, however, \textit{not} invalidate locality in the sense of ``Einstein causality'')!
These conclusions agree with ones reached by studying gedankenexperiments such as \textit{``Wigner's friend''} and other related ones, e.g., one recently proposed in \cite{Renner}.\\

Our task is thus to find out what one has to add to a minimal formulation of Quantum Mechanics in order to be able to describe the \textit{stochastic dynamics of states} of physical systems in the presence of ``events'' and their recordings (in projective measurements), in such a way that correlations between the outcomes of measurements agree with the Bell-type ``non-locality'' of Quantum Mechanics -- without the need to assume that ``observers'' intervene. The results reviewed in the next section are intended to report on some progress in this direction.

\section{Summary of the ``\textit{ETH}-Approach''}
In this section I briefly describe the so-called \textit{``ETH-Approach to Quantum Mechanics''} \cite{FS-Prob-Theory, BFS-forks, Schubnel-thesis, Les-Diablerets, FFS, FS-state-prep}, which is designed to retain attractive features of the Copenhagen Interpretation but eliminates its fatal weaknesses; and I note that ``$E$'' stands for ``Events'', ``$T$'' for ``Trees'', and ``$H$'' for ``Histories''. In the following, I attempt to explain what these terms mean, and why the concepts underlying the ``$ETH$-Approach'' are important for an understanding of the foundations of Quantum Mechanics ($QM$). The basic premises and contentions of this approach are as follows:

\begin{enumerate}
\item[I.]{\underline{Potential Events}. In the $ETH$-Approach to $QM$, \textit{Time}, denoted by $t$, is taken as an irreducible concept. It is described by the real line, $\mathbb{R}$, with its usual order relation.\footnote{The role of \textit{space-}time in a relativistic version of the ``$ETH$-Approach'' is discussed in \cite{Fr}} But in order to make the following discussion mathematically watertight it is advisable to sometimes assume that time is \textit{discretized}, $t \in \mathbb{Z}$. An important idea underlying the $ETH$-Approach is that time is not merely a parameter, but that it can be monitored by recording ``events'' happening in an isolated open system. (The precise meaning of this idea will become clearer later on.)
\\Let $t_0 \in \mathbb{R}$ be the time of the present. We consider an isolated open physical system $S$ and we denote by $\mathcal{H}$ the Hilbert space of pure state vectors of $S$. Our first task is to clarify what is meant by \textit{``potential events''} in $S$ that may happen at some future time $t>t_0$, or later: Potential events are described by families, $\lbrace \pi_{\xi}, \xi \in \mathcal{X} \rbrace$ of orthogonal projections acting on $\mathcal{H}$, with the properties that 
\begin{eqnarray}\label{pot-event}
\pi_{\xi}\cdot \pi_{\eta} &=& \delta_{\xi \eta}\, \pi_{\xi}, \,\,\forall \xi, \eta \,\,\text{in} \,\,\mathcal{X}, \quad(\text{disjointeness}) \nonumber\\
\sum_{\xi \in \mathcal{X}} \pi_{\xi}& =& {\mathds{1}}, \quad(\text{partition of unity}).
\end{eqnarray}
For simplicity we henceforth assume that the sets $\mathcal{X}$ labelling the projections that represent potential events are countable, discrete sets. (This merely serves to avoid technical complications in our exposition; of course, continuous spectra occur, too.)
In the Heisenberg picture, which we will use henceforth, the concrete projection operators acting on the Hilbert space 
$\mathcal{H}$ of $S$ representing a \textit{specific} potential event, e.g., the click of a detector belonging to $S$ when it is hit by a certain type of particle in $S$, depend on the time $t>t_{0}$ in the future when the event might happen. In an autonomous system, the concrete projection operators representing a \textit{specific} potential event that may happen at a time $t>t_0$ or at another time $t'>t_0$ are unitarily conjugated to one another by the propagator $U(t,t')$ of the system; (Heisenberg-picture evolution of operators). All projection operators representing potential events that may happen at some time $t>t_0$, or later, generate a $^{*}$-algebra denoted by $\mathcal{E}_{\geq t}$. It immediately follows from this definition that 
$$\mathcal{E}_{\geq t'} \subseteq \mathcal{E}_{\geq t}, \quad \text{if   }\, t'>t.$$

\underline{\bf{Remark}}: The concrete projection operators representing some potential event that may happen in system $S$ (see Eqs. \eqref{pot-event}) depend
on the time $t$ when the potential event would start to happen and on the time-interval during which it would happen. More concretely, if $\hat{A}_{i}\,, i=1,2, \dots,$ are abstract operators representing physical quantities of $S$, (e.g., a component of the spin of a certain species of particles localized in a certain region of physical space and measured in a Stern-Gerlach experiment), and if $A_{i}(t)$ denotes the Heisenberg-picture operator on $\mathcal{H}$ representing $\hat{A}_i$ at time $t$, then a potential event arising from monitoring the quantities 
$\hat{A}_i \,,i=1,2, \dots,$ which starts to happen at time $t$, consists \,of a \,family of projections
satisfying Eqs. \eqref{pot-event} that are functionals of the operators 
$$\hspace{0.5cm} \lbrace A_{i}(t') \vert\, i=1,2, \dots;\, t' \in [\,t, T),\,\text{for some }\,T\, \text{ with  }\, t<T \leq \infty \rbrace \hspace{0.7cm} \square$$
This remark is inspired by general wisdom from local quantum field theory.\\

For simplicity we assume that \textit{all physically relevant states} of $S$ can be described by density matrices acting on $\mathcal{H}$, and that the algebras $\mathcal{E}_{\geq t}$ are closed in the weak topology of the algebra, $B(\mathcal{H})$, of all bounded operators acting on $\mathcal{H}$. Typically, all the algebras $\mathcal{E}_{\geq t}$ are then isomorphic to one \textit{universal} (von Neumann) algebra\footnote{In local relativistic quantum theories with massless particles, the algebra $\mathcal{N}$ tends to be a von Neumann algebra of type $III$; see \cite{Buchholz}} $\mathcal{N}$, i.e., 
\begin{equation}\label{iso}
\mathcal{E}_{\geq t} \simeq \mathcal{N}, \quad \forall t \in \mathbb{R}.
\end{equation}
The algebra, $\mathcal{E}$, of all potential events that may happen in the course of history is defined by
\begin{equation}\label{allevents}
B(\mathcal{H}) \supseteq \mathcal{E}:= \overline{\bigvee_{t \in \mathbb{R}} \mathcal{E}_{\geq t}}\,,
\end{equation}
(where the closure is taken in the operator norm of $B(\mathcal{H})$).
}\\

\item[II.]{\underline{The Principle of Diminishing Potentialities}. In the quantum theory of (autono-\\mous) systems with finitely many degrees of freedom -- as treated in our introductory courses on $QM$ -- the algebras $\mathcal{E}_{\geq t}$ turn out to be \textit{independent} of time $t$; and usually $\mathcal{E}_{\geq t}= B(\mathcal{H})$. For such systems, one \textit{cannot} develop a sensible quantum theory of events, and it is impossible to come up with a logically coherent, intrinsically quantum-mechanical description of the retrieval of information on such systems, i.e., of measurements, \textit{without} adding further quantum systems with infinitely many degrees of freedom that serve to ``measure'' the former systems; (or without resorting to something like ``Copenhagen''). In this respect, quantum systems with finitely many degrees of freedom are as ``interesting'' as the space-time region outside the event horizon of a black hole: no information can be extracted! In order to encounter non-trivial dependence of the algebras $\mathcal{E}_{\geq t}$ on time $t$, we must consider \textit{isolated (open) systems with infinitely many degrees of freedom} and with the property that the propagator $U$ of $S$ is generated by a Hamiltonian whose spectrum does \textit{not} have any isolated eigenvalues, and (if time is continuous) the spectrum is \textit{unbounded above and below}, or, in relativistic quantum theory, it is \textit{semi-bounded, but without any spectral gaps}; i.e., we must assume that there exist massless modes.

 Our contention is that a basic property of a quantum theory of isolated open systems, $S$, enabling one to describe \textit{events} and their \textit{recording} in projective measurements of physical quantities is captured in the following {\bf{``Principle of Diminishing Potentialities''}} ($PDP$):
\begin{equation}\label{PDP}
\boxed{\mathcal{E}_{\geq t'} \subsetneqq \mathcal{E}_{\geq t} \subsetneqq \mathcal{E}, \quad \text{whenever}\,\,\,\, t'>t.}
\end{equation}
To be more precise, one expects that if time is continuous the relative commutant
$$\big(\mathcal{E}_{\geq t'}\big){'} \cap \mathcal{E}_{\geq t},\quad \text{with   }\,\, t'>t,$$
is an infinite-dimensional, non-commutative algebra. (If time is discrete this relative commutant can, however, be a finite-dimensional algebra.) Examples of non-relativistic and relativistic systems satisfying property \eqref{PDP} will be discussed elsewhere, (see also \cite{Les-Diablerets}).\footnote{I sometimes fear that unrealistically simple examples advanced with the intention to clarify aspects of the foundations of $QM$ have had the opposite effect: They have contributed to clouding our views.} Here I just mention that ($PDP$), in the sense of a relativistic variant of Eq. \eqref{PDP}, is a \textit{theorem} in local relativistic quantum field theories with massless particles in four space-time dimensions.\footnote{and the algebras $\mathcal{E}_{\geq t}, t \in \mathbb{R}$, are von Neumann algebras of type $III$.} This follows from important results in \cite{Buchholz} and is used in \cite{Fr}.

\underline{\bf{Definition 1}}. \textit{Isolated open systems} $S$ (featuring events) are henceforth \textit{defined} in terms of a filtration, $\lbrace \mathcal{E}_{\geq t} \rbrace_{t \in \mathbb{R}}$ (or, for the sake of simplicity and precision, $\lbrace \mathcal{E}_{\geq t} \rbrace_{t \in \mathbb{Z}}$), of (von Neumann) algebras satisfying the \,\textit{``Principle of Diminishing Potentialities''} \eqref{PDP}, all represented on a common Hilbert space $\mathcal{H}$, whose projections describe potential events. \hspace{4.5cm} $\square$\\

If $\Omega$ denotes the density matrix on $\mathcal{H}$ representing the actual state of a system $S$ we use the notation
$$\omega(X):= tr(\Omega\,X), \qquad \forall X \in B(\mathcal{H}),$$
to denote the expectation value of the operator $X$ in the state $\omega$ determined by $\Omega$. We define
\begin{equation}\label{state-rest}
\omega_{\,t}(X):= \omega(X), \qquad \forall X \in \mathcal{E}_{\geq t},
\end{equation}
i.e., $\omega_{\,t}$ is the \textit{restriction} of the state $\omega$ to the algebra $\mathcal{E}_{\geq t}$.\\
Note that, as a consequence of ($PDP$) and of \textit{entanglement}, the restriction, $\omega_{\,t}$, of a state $\omega$ on the algebra
$\mathcal{E}$ to a subalgebra $\mathcal{E}_{\geq t} \subset \mathcal{E}$ will usually be {\bf{mixed}} \textit{even} if $\omega$ is a {\bf{pure}} state on $\mathcal{E}$.
}\\

\item[III.]{\underline{Actual Events}. 
Henceforth we only study isolated open systems $S$ for which ($PDP$), in the form of Eq. \eqref{PDP}, holds. Let \mbox{$\lbrace \pi_{\xi}, \xi \in \mathcal{X} \rbrace \subset \mathcal{E}_{\geq t}$} be a potential event that might start to happen at some time $t$, with  \mbox{$\lbrace \pi_{\xi}, \xi \in \mathcal{X} \rbrace$} \textit{not} contained in $\mathcal{E}_{\geq t'}, \,\text{  for   }\, t'>t$. Tentatively, we say that this potential event {\bf{actually starts to happen}} at time $t$ iff
\begin{equation}\label{inc-superpos}
\omega_{\,t}(X)=\sum_{\xi \in \mathcal{X}} \omega_{\,t}\big( \pi_{\xi} \,X\, \pi_{\xi} \big), \qquad \forall X \in \mathcal{E}_{\geq t},
\end{equation}
meaning that $\omega_{\,t}$ is an incoherent superposition of states labelled by the points $\xi \in \mathcal{X}$; in other words, off-diagonal expectations, 
$\omega_{\,t}\big( \pi_{\xi}\,X\, \pi_{\eta} \big), \, \xi \not= \eta,$ do \textit{not} contribute to the right side of \eqref{inc-superpos}.
Equation \eqref{inc-superpos} is equivalent to saying that the projections $\pi_{\xi}, \xi \in \mathcal{X},$ belong to the \textit{centralizer} of the state $\omega_t$.

 Given a $^{*}$-algebra $\mathcal{M}$ and a state $\omega$ on $\mathcal{M}$, the centralizer, $\mathcal{C}_{\omega}(\mathcal{M})$, of the state $\omega$ is defined to be the subalgebra of $\mathcal{M}$ spanned by all operators, $Y$, in $\mathcal{M}$ with the property that 
$$\omega([Y, X]) =0, \qquad \forall X \in \mathcal{M}.$$
The \textit{center} of the centralizer, denoted by $\mathcal{Z}_{\omega}(\mathcal{M})$, is the abelian subalgebra of the centralizer  consisting of all operators in $\mathcal{C}_{\omega}(\mathcal{M})$ commuting with all other operators in 
$\mathcal{C}_{\omega}(\mathcal{M})$.\\
 We note that the center, $\mathcal{Z}(\mathcal{M})$, of the algebra $\mathcal{M}$ is contained in 
 $\mathcal{Z}_{\omega}(\mathcal{M})$, \textit{for all} states $\omega$.\\

\underline{\bf{Definition 2}}. A potential event $\lbrace \pi_{\xi}, \xi \in \mathcal{X} \rbrace \subset \mathcal{E}_{\geq t}$, with 
$\lbrace \pi_{\xi}, \xi \in \mathcal{X} \rbrace$ not contained in $\mathcal{E}_{\geq t'}, \,\text{  for  }\, t'>t$, 
\textit{actually starts to happen} at time $t$ iff $\mathcal{Z}_{\omega_t}(\mathcal{E}_{\geq t})$ is \textit{non-trivial},
\begin{equation}\label{act-event}
\lbrace \pi_{\xi}, \xi \in \mathcal{X} \rbrace \,\,\text{generates   }\,\, \mathcal{Z}_{\omega_{\,t}}\big(\mathcal{E}_{\geq t}\big),
\end{equation}
and 
\begin{equation}\label{real-alt}
\omega_{\,t}(\pi_{\xi_{j}}) \,\,\text{ is \textit{strictly positive}},\, \,\, \xi_j \in \mathcal{X}, \,\, j=1,2, \dots, n\,, 
\end{equation}
for some \, $n\geq 2$. \hspace{7cm} $ \square$
}\\

\item[IV.]{\underline{The fundamental Axiom}. We are now in a position to describe the evolution of states in the $ETH$-Approach to QM. Let $\omega_{\,t}$ be the state of an isolated system $S$ right before time $t$. Let us suppose that an event $\lbrace \pi_{\xi}, \xi \in \mathcal{X} \rbrace$ generating $\mathcal{Z}_{\omega_t}(\mathcal{E}_{\geq t})$ starts to happen at time $t$, in the sense of Definition 2. \\

\underline{\bf{Axiom}}. The actual state of the system $S$ right \textit{after} time $t$ when the event $\lbrace \pi_{\xi}, \xi \in \mathcal{X} \rbrace$ has started to happen is given by one of the states
\begin{equation}\label{SCP}
\omega_{\,t, \xi_{*}}(\cdot):=[\omega_{\,t}(\pi_{\xi_{*}})]^{-1}\,\omega_{\,t}\big(\pi_{\xi_{*}} (\cdot) \pi_{\xi_{*}}\big)\,,
\end{equation}
for some $\xi_{*} \in \mathcal{X}$ with $\omega_{\,t}(\pi_{\xi_{*}})>0$, (``{\bf{state-collapse}} postulate''\footnote{a rather unfortunate name!}). The probability for the system $S$ to be found in the state $\omega_{\,t,\xi_{*}}$ right after time $t$ when the event 
$\lbrace \pi_{\xi}, \xi \in \mathcal{X} \rbrace$ has started to happen is given by {\bf{Born's Rule}}, i.e., by
\begin{equation}\label{Born}
\hspace{4.1cm}prob\{\xi_{*}, t\} = \omega_{\,t}(\pi_{\xi_{*}}). \hspace{2.2cm} \square
\end{equation}
\textit{Remarks:} \\
\textit{(1)} The projection $\pi_{\xi_{*}}$ selecting the actual state $\omega_{\,t, \xi_{*}}$ of $S$ (and sometimes also the point 
$\xi_{*} \in \mathcal{X}$) is called the \textit{``actual event''} happening at time $t$. \\
\textit{(2)} The contents and meaning of this Axiom are clear and mathematically watertight as long as time is discrete. (If time is continuous further precision ought to be provided.)}
\end{enumerate}
This Axiom, Eqs. \eqref{SCP} and \eqref{Born}, conveys the following picture of quantum dynamics: In Quantum Mechanics, the \underline{evolution of states} of an isolated open system $S$ featuring events, in the sense of Definitions 1 and 2 proposed above, is given by a (rather unusual novel type of) \underline{stochastic branching process}, whose state space is what I call the \textit{``non-commutative spectrum''}, 
$\mathfrak{Z}_{S}$, of $S$. Assuming that Eq. \eqref{iso} holds, the non-commutative spectrum of $S$ is defined by
\begin{equation}\label{NCspect}
\mathfrak{Z}_{S}:= \bigcup_{\omega} \mathcal{Z}_{\omega}(\mathcal{N})\,, \quad \text{with} \quad \mathfrak{X}_{S}:= \bigcup_{\omega} \text{spec}\Big(\mathcal{Z}_{\omega}(\mathcal{N})\Big),
\end{equation}
where the union over $\omega$ is a disjoint union, and $\omega$ ranges over \textit{all} physical states of $S$.\footnote{The set $\mathfrak{X}_{S}$ can also be defined in terms of a certain ``flag manifold'' associated with the Hilbert space $\mathcal{H}$} Eq. \eqref{act-event} and {\bf{Born's Rule}}, Eq. \eqref{Born}, specify the \textit{branching probabilities} of the process.\\
\begin{center}
\includegraphics[height=8.4cm]{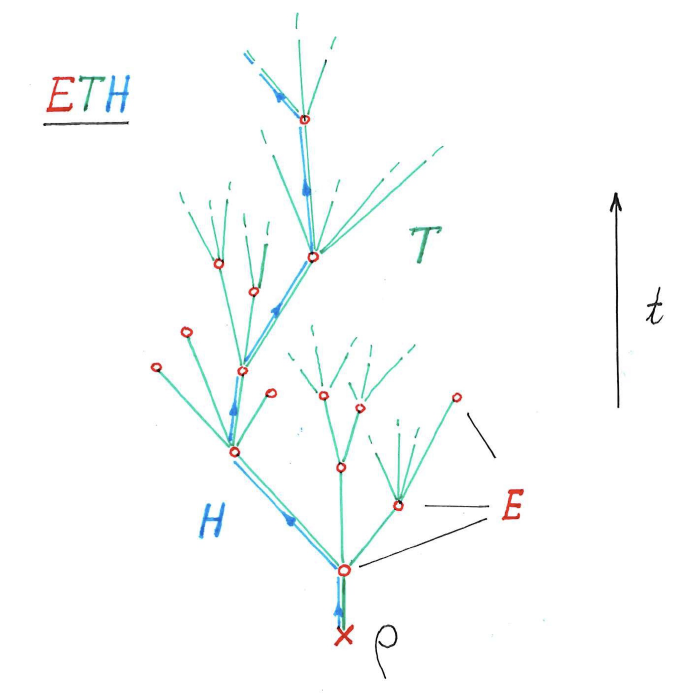}\\
Time evolution of a state of $S$ with initial condition $\omega:= \rho$\\
{\color{red}$E$}: ``Events'', {\color{caribbeangreen}$T$}: ``Tree'' of possible future states, {\color{blue}$H$}: ``History'' of actual events/states.\\ 
\textit{Figure 3}
\end{center}
The above picture of the stochastic time evolution of states of an isolated open system $S$ is illustrated, metaphorically (for discrete time), in Figure 3. It differs substantially from and supercedes the \textit{``decoherence mumbo-jumbo''}.

Let us suppose, for the sake of simplicity and mathematical precision, that time is discrete, ($t\in \mathbb{Z}$). It is important to note that, in general, the events (described by orthogonal projections in $\mathcal{E}_{\geq t'}$) predicted to happen at a later time $t'>t$ on the basis of the states $\omega_{t, \xi}, \xi \in \mathcal{X}$, where \mbox{$\lbrace \pi_{\xi}, \xi \in \mathcal{X} \rbrace$} generates $\mathcal{Z}_{\omega_t}(\mathcal{E}_{\geq t})$, are \textit{different} from the events one would predict to happen at time $t'$ on the basis of the state $\omega_{t}\vert_{\mathcal{E}_{\geq t'}}$, used when the actual event happening at time $t$ is not known (i.e., has not been recorded); and the projections representing these different sets of events usually do \textit{not} commute with one another. 
Furthermore, for $t'>t$, the operators in $\mathcal{Z}_{\omega_{t, \xi}}(\mathcal{E}_{\geq t'})$ and in 
$\mathcal{Z}_{\omega_{t, \eta}}(\mathcal{E}_{\geq t'}), \xi, \eta \in \mathcal{X},$ (with $\omega_{t}(\pi_{\xi}), \,\omega_{t}(\pi_{\eta})$ strictly positive), but $\xi \not= \eta$, do \textit{not} in general commute with each other. This is a \textit{fundamental difference} between the \textit{``non-commutative branching processes''}, described here, and classical stochastic branching processes.\\
The discussion above is mathematically sound if time is discrete, but requires more precision if time is taken to be continuous.\\
To be on the safe side, we temporarily choose time to be discrete ($t\in \mathbb{Z}$). Let $H$ be the Hamiltonian of an isolated open system, and suppose that 
\begin{equation} \label{H}
\Vert e^{iH} - {\bf{1}} \Vert \ll 1\,.
\end{equation}
Let us suppose that $\lbrace \pi_{t,\xi} , \xi \in \mathcal{X}_{t} \rbrace$ is an event that starts to happen at time $t$, provided the state of $S$ at time $t$ is given by $\omega_{\,t}$; (i.e., $\lbrace \pi_{t,\xi} , \xi \in \mathcal{X}_{t} \rbrace$ generates $\mathcal{Z}_{\omega_{\,t}}(\mathcal{E}_{\geq t})$). Let $\xi_{*}$ be the element of $\mathcal{X}_{t}$ with the property that, in accordance with the Axiom stated in IV., above, the state of $S$ right after time $t$ is given by 
$$\omega_{\,t, \xi_{*}}(\cdot):=[\omega_{\,t}(\pi_{t, \xi_{*}})]^{-1}\,\omega_{\,t}\big(\pi_{t, \xi_{*}} (\cdot) \pi_{t, \xi_{*}}\big)\,,$$
with $\omega_{\,t}\big( \pi_{t, \xi_{*}} \big) > 0$; i.e., $\pi_{t,\xi_{*}}$ is the ``actual event'' happening at time $t$. Let $t' = t+1$ be the time following $t$, and let 
\mbox{$\lbrace \pi_{t', \xi}, \xi \in \mathcal{X}_{t'} \rbrace$} be the event that starts to happen at time $t'$, \textit{provided} that the state of $S$ at time $t'$ is given by $\omega_{\,t, \xi_{*}}$. Then assumption \eqref{H} suggests that there exists an element 
$\xi_{\natural}\in \mathcal{X}_{t'}$ with the property that 
\begin{eqnarray} \label{Continuity}
\omega_{\,t, \xi_{*}}\big(\pi_{t', \xi_{\natural}}\big) &\approx & 1, \,\,\text{ but }\nonumber \\
\omega_{\,t, \xi_{*}}\big(\,\pi_{t', \xi}\,\big) &\ll& 1, \,\,\forall\,\, \xi \not= \xi_{\natural}\,,\,\xi \in \mathcal{X}_{t'}\,.
\end{eqnarray}
According to the Axiom in IV., in particular {\bf{Born's Rule}}, the actual state of $S$ right after time $t'$ is then very likely given by 
$$\omega_{\,t, \xi_{*}, t',\xi_{\natural}}(\cdot) := [\omega_{\,t, \xi_{*}}(\pi_{t', \xi_{\natural}})]^{-1}\omega_{\,t, \xi_{*}}\big(\pi_{t', \xi_{\natural}} (\cdot) \pi_{t', \xi_{\natural}}\big) \approx \omega_{\,t, \xi_{*}}(\cdot)\,.$$
The state $\omega_{\,t, \xi_{*}, t',\xi_{\natural}}$ is close to the one that would commonly be used in the Heisenberg picture of quantum mechanics in the absence of any ``measurements'' or ``events'' after time $t$, namely the state 
$\omega_{\,t, \xi_{*}}(\cdot)$.

However, for purely statistical (\textit{entropic}!) reasons, every once in a while, i.e., at rare times $t'$, an event $\pi_{t',\xi}$ is realised that has a \textit{very small Born probability}, $\omega_{\,t'}(\pi_{t',\xi}) \ll 1,    \quad \xi \in \mathcal{X}_{t'}\,.$\\

\textit{Digression on ``Missing Information'' associated with events:}\footnote{This digression can be omitted at first reading, and the reader is invited to proceed to point V., below.}

Given the event $\lbrace \pi_{t,\xi}, \xi \in \mathcal{X}_{t} \rbrace$ happening at time $t$, assuming that $\omega_{t}$ is the actual state of $S$ right before time $t$, we define the \textit{``missing information''} (or \textit{``entropy production''}\,), $\sigma(\omega_{t}, \mathcal{X}_{t})$, associated with this event by
\begin{equation}\label{entropy}
\sigma(\omega_{t}, \mathcal{X}_{t}):= - \sum_{\xi \in \mathcal{X}_{t}} \omega_{t}(\pi_{t,\xi})\cdot \ell n\big( \omega_{t}(\pi_{t,\xi})\big)
\end{equation}
Assuming that \eqref{H} holds, the ``missing information'' associated with most events that ever happen is very small. If the ``missing information'' associated with \textit{all} events were tiny then taking the state of $S$ in the Heisenberg picture to be constant in time would be a good approximation to its stochastic evolution. However, every once in a while, events corresponding to a \textit{large} ``missing information'' (entropy production) may be encountered, and these are the events that will most likely be noticed and recorded, because they trigger a substantial change of the state of $S$. (Some people will want to call them ``measurements''.)

Let $t_0$ be the time at which the system $S$ has been prepared in a state $\omega$, (as discussed in \cite{FS-state-prep}), and $t_j:=t_0 + j \in \mathbb{Z}$; further, let $\pi_{t_j, \xi_j}$ be the \textit{actual event} happening at time $t_j$, given the initial state $\omega$ of $S$ and earlier actual events $\pi_{t_\ell, \xi_{\ell}}, \ell < j,$ $j=1,2, \dots, n$; (see Definition 2 and Axiom). We define 
\begin{equation}\label{Mu_omega}
\mu_{\omega}\big(\xi_1, \xi_2, \dots, \xi_n\vert X\big):= \omega\Big(\prod_{j=1}^{n} \pi_{t_j, \xi_j}\cdot X\cdot X^{*} \cdot (\prod_{j=1}^{n}\pi_{t_j, \xi_j})^{*}\Big)\,,
\end{equation}
where the product is ordered according to $\prod_{j=1}^{n} a_{j} = a_1 \cdot a_2 \cdots a_n$, and $X$ is an arbitrary non-zero operator in $\mathcal{E}_{\geq t}$, for some $t> t_n$, with $\omega\big(X\cdot X^{*}\big)>0$.
Then $\mu_{\omega}(\dots \vert X)$ is a positive measure on the Cartesian product ${\bigtimes}_{j=1}^{n} \mathcal{X}_{t_j}$. \textit{Note that the space $\mathcal{X}_{t_{k+1}}$ depends on the choice of $\omega$ and on {\bf{all}} the actual events $\pi_{t_1,\xi_1}, \dots, \pi_{t_{k},\xi_{k}}$ that happened at times $t_1 <\dots < t_k$, {\bf{before}} $t_{k+1}$; with $k=1,2, \dots, n-1$.} For any $m$, with $0<m<n$, we set
$$X(\underline{\xi}^{(m,n)}):= \prod_{j=m+1}^{n} \pi_{t_j, \xi_j} \cdot X\,,$$
and $X(\underline{\xi}^{(n,n)}):=X$. Then 
$$\mu_{\omega}\big(\xi_1,\dots,\xi_n \vert X\big)= \mu_{\omega}\big(\xi_1, \dots, \xi_m \vert X(\underline{\xi}^{(m,n)})\big)\,.$$
The measure 
$\mu_{\omega}\big(\dots \vert X\big)$ has the (possibly somewhat perplexing) property that\\

\quad $\underset{\xi_{k+1},\dots, \xi_{m}}{\sum} \mu_{\omega} \big(\xi_1,\dots, \xi_{k}, \xi_{k+1}, \dots, \xi_{m}\vert X(\underline{\xi}^{(m,n)})\big) = $
\begin{equation}\label{sumrule}
\hspace{4cm} = \mu_{\omega}(\xi_1, \dots, \xi_{k} \vert X(\underline{\xi}^{(m,n)})\big)\,,
\end{equation}
for arbitrary $k$, with $1\leq k \leq m\leq n$, as one easily verifies. (Identity \eqref{sumrule} may look familiar to the reader from a similar one satisfied by the ``L\"{u}ders-Schwinger-Wigner formula'' \cite{Schwinger} for the probability of a sequence of outcomes of measurements, assuming perfect \textit{decoherence}. However, it actually has quite a different origin!) It is sometimes convenient to define $\mu_{\omega}\big(\dots \vert X\big)$ as a measure on the space
$$\mathfrak{X}_{n}:= \big(\mathfrak{X}_{S}\big)^{\times n}\,,$$
where $\mathfrak{X}_{S}$ has been defined in Eq. \eqref{NCspect}, with the convention that 
$$\pi_{t_{k}, \xi}=0, \,\,\text{  unless  }\,\, \xi \in \mathcal{X}_{t_k} \subset \mathfrak{X}_{S}\,.$$
For $X ={\mathds{1}}$, $\mu_{\omega}(\dots\vert {\mathds{1}})$ is a \textit{probability measure} on $\mathfrak{X}_{n}$. If arbitrarily long sequences of events are considered it is useful to introduce the ``path space''
$$\mathfrak{X}_{\infty}:=\underset{n \rightarrow \infty}{\underrightarrow{\text{lim}}} \mathfrak{X}_{n}\,.$$
Thanks to property \eqref{sumrule}, the measures $\mu_{\omega}(\dots \vert {\mathds{1}})$ determine a unique probability measure on $\mathfrak{X}_{\infty}$. This follows from a well known lemma due to \textit{Kolmogorov}.\\
Next, we define the \textit{``missing information per event''} of a sequence of events, as follows:
$$\sigma_{n}(\mu_{\omega}):= - \frac{1}{n}\sum_{\xi_1, \dots, \xi_n} \mu_{\omega}(\xi_1, \dots, \xi_n\vert {\mathds{1}}) 
\cdot \ell n\big(\mu_{\omega}(\xi_1, \dots, \xi_n \vert {\mathds{1}})\big)\,, $$
and 
\begin{equation}\label{spec-entropy}
\sigma(\mu_{\omega}):= \text{limsup}_{n\rightarrow \infty} \sigma_{n}(\mu_{\omega})
\end{equation}
If events happening at times $t_1,\dots, t_n$ are not recorded then $\sigma_{n}(\mu_{\omega})$ is a measure of how much the state of the system at time $t>t_n$ deviates from the (initial) state $\omega$ used in the Heisenberg picture of standard $QM$.

Of particular interest is the so-called \textit{relative entropy} 
\begin{eqnarray}\label{relentropy}
S_{n}\big(\mu_{\omega} \Vert \mu_{\omega}^{opp}\big)&:= &\sum_{\xi_1, \dots, \xi_{n}} \mu_{\omega}(\xi_1, \dots, \xi_{n}\vert {\mathds{1}}) \times \nonumber \\
&\times& \Big(\ell n  \,\mu_{\omega}(\xi_1, \dots, \xi_{n}\vert {\bf{1}}) - \ell n \, \mu_{\omega}^{opp}(\xi_1, \dots, \xi_{n}\vert {\mathds{1}}) \Big)\,,
\end{eqnarray}
where
$$\mu_{\omega}^{opp}(\xi_1, \dots, \xi_n \vert {\mathds{1}}) := \omega\Big((\prod_{j=1}^{n} \pi_{t_j, \xi_j})^{*}\cdot \prod_{j=1}^{n}\pi_{t_j, \xi_j}\Big)\,$$
is the measure obtained when the order of the events is (time-)reversed. The relative entropy $S_{n}\big(\mu_{\omega} \Vert \mu_{\omega}^{opp}\big)$ is \textit{non-negative}, and its growth in $n$, as $n \rightarrow \infty$, is a measure of the \textit{irreversibility} of histories of events featured by the system and reflects the ``arrow of time''.

 \textit{End of Digression.}

\begin{enumerate}
\item[V.]{\underline{Recording events by ``projective measurements'' of physical quantities}. We consider an isolated open system $S$ described in terms of a filtration $\lbrace \mathcal{E}_{\geq t} \rbrace_{t \in \mathbb{R}}$ of algebras represented on its Hilbert space $\mathcal{H}$ of pure state vectors, as described in Defnition 1, (paragraph I.). We propose to clarify how events happening in $S$ can be recorded by projectively (directly) measuring \textit{``physical quantities''} characteristic of $S$. (Time may be taken to be continuous; but, for the sake of simplicity and mathematical precison, the reader is invited to continue to assume that $t \in \mathbb{Z}$.) \\

\underline{\bf{Definition 3}}. A \textit{``physical quantitiy''} characteristic of $S$ is an abelian ($C^{*}$-) algebra, $\mathcal{Q}$, with the property that, for each time $t$, there exists a representation, $\sigma_{t}^{\mathcal{Q}}$, of $\mathcal{Q}$ on $\mathcal{H}$ as a subalgebra of $\mathcal{E}_{\geq t}$. \hspace{3.5cm} $\square$\\

For autonomous systems, the representations $\sigma_{t}^{\mathcal{Q}}$ and $\sigma_{t'}^{\mathcal{Q}}$ are unitarily equivalent, with
$$\sigma_{t}^{\mathcal{Q}}(A) = U(t',t)\,\sigma_{t'}^{\mathcal{Q}}(A)\, U(t,t'), \quad \forall A \in \mathcal{Q}\,,$$
where $U(t',t)= \text{exp}\big(i(t-t')H\big)$ is the propagator of $S$, with $t,t'$ arbitrary times; (Heisenberg-picture dynamics).

For simplicity, we assume that the physical quantities $\mathcal{Q}$ available to identify properties of $S$ or record events all have discrete spectrum; i.e., 
\begin{equation}\label{phys-Q}
\mathcal{Q} = \langle \Pi_{\eta}^{\mathcal{Q}} \vert \eta \in \mathcal{Y}^{\mathcal{Q}} \rangle,
\end{equation}
where $\mathcal{Y}^{\mathcal{Q}} \equiv \text{spec}(\mathcal{Q})$ is a discrete set, which we view as a subset of the real line, and the operators $\Pi_{\eta}^{\mathcal{Q}}$ are disjoint orthogonal projections. (Of course, continuous spectra can arise, too. But in order to avoid technical complications, we ignore them here.)
We can then describe $\mathcal{Q}$ as the algebra given by all functions of a single self-adjoint operator, $\widehat{Y}$, with
discrete spectrum, spec$(\widehat{Y}) \simeq \mathcal{Y}^{\mathcal{Q}}$, and spectral projections $\Pi_{\eta}^{\mathcal{Q}}$. For every time $t$, there exists a self-adjoint operator,
$Y(t)=\sigma_{t}^{\mathcal{Q}}(\widehat{Y})$, acting on $\mathcal{H}$  that represents $\widehat{Y}$ at time $t$. \\
It is interesting to ask whether physical quantities can serve to detect or record events happening in $S$. For a discrete set
$$\mathcal{O}_{S}= \lbrace \mathcal{Q}_j \rbrace_{j \in \mathfrak{J}}$$
 of physical quantities characteristic of $S$, it is arbitrarily unlikely that one of the algebras $\sigma_{t}^{\mathcal{Q_{\textit{j}}}}(\mathcal{Q}_{j} )$,\,$j\in \mathfrak{J},$ has a non-trivial intersection with (e.g., contains or is contained in) an algebra 
 $\mathcal{Z}_{\omega_{\,t}}(\mathcal{E}_{\geq t})$ describing the event happening at time $t$, for some state $\omega_{\,t}$. To cope with this problem, we have to understand how well $\mathcal{Z}_{\omega_{\,t}}(\mathcal{E}_{\geq t})$ can be approximated by an algebra generated by a family, $\lbrace Q_{\alpha}(t) \rbrace_{\alpha=0}^{N}$, of disjoint orthogonal projections  contained in (or equal to) an algebra $\sigma_{t}^{\mathcal{Q}}(\mathcal{Q})$, for some $\mathcal{Q} \in \mathcal{O}_{S}$.
 
  There are different ways of quantifying how well the algebra generated by $\lbrace Q_{\alpha}(t) \rbrace_{\alpha=0}^{N}$ approximates the event described by $\mathcal{Z}_{\omega_t}(\mathcal{E}_{\geq t})$. To keep our discussion brief, it is 
 convenient to introduce \textit{``conditional expectations''} of algebras:\\
 
 \underline{\bf{Definition 4}}. \\
 Let $\mathcal{N}$ be a (von Neumann) subalgebra of a (von Neumann) algebra $\mathcal{M}$. A linear map 
 \begin{equation} \label{condexp}
 \epsilon_{\omega} : \mathcal{M} \underset{\text{onto}}{\rightarrow} \mathcal{N}
 \end{equation}
 is a \textit{conditional expectation} from $\mathcal{M}$ onto $\mathcal{N}$ with respect to a normal state $\omega$ on 
 $\mathcal{M}$ iff
\begin{enumerate}
\item[(i)]{$\Vert \epsilon_{\omega} (X) \Vert \leq \Vert X \Vert,$ \quad $\forall X \in \mathcal{M}$}
\item[(ii)]{ $\epsilon_{\omega}(X) = X$, \quad $\forall X \in \mathcal{N}$}
\item[(iii)]{ $\omega \circ \epsilon_{\omega} = \omega$} 
\item[(iv)]{$\epsilon_{\omega}(AXB)= A \epsilon_{\omega}(X) B, \quad \forall \, A, B, \in \mathcal{N}, \, \forall \, X \in \mathcal{M}$}\hspace{2.8cm}$\square$
 \end{enumerate}

Conditional expectations have the following properties:
\begin{enumerate}
\item[(v)]{$ \epsilon_{\omega}(X^{*}X) \geq 0, \quad \forall \,X \in \mathcal{M}$}
\item[(vi)]{$\epsilon_{\omega} : \mathcal{M} \rightarrow \mathcal{N}$ is completely positive, and 
$\epsilon_{\omega}({\mathds{1}}_{\mathcal{M}})= {\mathds{1}}_{\mathcal{N}}$}
\end{enumerate}
 See, e.g., \cite{Takesaki} for an exposition of the theory of conditional expectations. Under very general assumptions, there exist conditional expectations
 \begin{equation}\label{condexp}
\epsilon_{\omega_{\,t}}: \mathcal{E}_{\geq t} \rightarrow \mathcal{Z}_{\omega_{\,t}}\big(\mathcal{E}_{\geq t}\big)\,,
\end{equation}
 for arbitrary times $t$.
 
Let $\omega_{\,t}$ be the state of a system $S$ right before an event $\lbrace \pi_{\xi}, \xi \in \mathcal{X}_{t}\rbrace$ generating 
$\mathcal{Z}_{\omega_{\,t}}(\mathcal{E}_{\geq t})$ starts to happen. I propose to clarify in which way a physical quantity 
$\mathcal{Q} \in \mathcal{O}_{S}$ can be used to record this event, and how precisely the value of this quantity identifies the \textit{actual} event, $\xi_{*} \in \mathcal{X}_t$, happening at time $t$.\\
We assume that there exists a physical quantity $\mathcal{Q}$ and a family of disjoint orthogonal projections $\lbrace \widehat{Q}_{\alpha} \rbrace_{\alpha=0}^{N} \subset \mathcal{Q},\, N\geq 2,$ with the following properties:
\begin{enumerate}
\item[(a)]{$\sum_{\alpha=0}^{N} Q_{\alpha}(t) = \mathds{1}$, \, where\, $Q_{\alpha}(t) = \sigma_{t}^{\mathcal{Q}}(\widehat{Q}_{\alpha}),\,
\alpha = 1,\dots, N, \,\, \forall t$;}
\item[(b)]{there exists a positive number $\delta\ll 1$ such that 
$$\omega_{\,t}\Big(\sum_{\alpha=1}^{N}Q_{\alpha}(t)\Big) \geq 1-\delta \quad (\text{or, equivalently,} \,\,\, \omega_{\,t}\big(Q_{0}(t)\big) \leq \delta\,); $$}
\item[(c)]{Given an operator $X\in \mathcal{E}_{\geq t}$, we define 
$$\text{dist}\big(X, \mathcal{Z}_{\omega_{\,t}}(\mathcal{E}_{\geq t})\big):= \Vert X - \epsilon_{\omega_{\,t}} (X) \Vert.$$
We assume that
\begin{equation}\label{dist}
\text{dist} \big(Q_{\alpha}(t), \mathcal{Z}_{\omega_{\,t}}(\mathcal{E}_{\geq t})\big)  < \delta\,, \quad \text{for }\,\, \alpha=1,\dots, N\,.
\end{equation}
}
\end{enumerate}
In the following, we use the notation $\mathcal{O}(\varepsilon)$ to denote any real number whose absolute value is bounded above by
$const.\, \varepsilon$, where $const.$ is a \textit{uniformly bounded} positive constant. Properties (a) through (c) of 
$\lbrace \widehat{Q}_{\alpha} \rbrace_{\alpha=0}^{N}$ can be used to derive the following equations:

For an arbitrary operator $X\in \mathcal{E}_{\geq t}$,
\begin{eqnarray}\label{superpos}
\omega_{\,t}(X)&=& \sum_{\alpha=1}^{N} \omega_{\,t} \big(Q_{\alpha}(t)\,X\big) + \mathcal{O}(\delta \Vert X \Vert) \nonumber\\
&=& \sum_{\alpha=1}^{N} \omega_{\,t}\big(Q_{\alpha}(t)[Q_{\alpha}(t) X]\big) + \mathcal{O}(\delta \Vert X \Vert) \nonumber\\
&=& \sum_{\alpha=1}^{N} \omega_{\,t}\big(\epsilon_{\omega_{\,t}}(Q_{\alpha}(t))[Q_{\alpha}(t) X]\big) + \mathcal{O}(\delta \,N \Vert X \Vert) \nonumber \\
\quad &=& \sum_{\alpha=1}^{N} \omega_{\,t}\big(Q_{\alpha}(t) X\, \epsilon_{\omega_{\,t}}(Q_{\alpha}(t))\big) + \mathcal{O}(\delta \,N \Vert X \Vert) \nonumber\\
&=&  \sum_{\alpha=1}^{N} \omega_{\,t}\big(Q_{\alpha}(t) X\,Q_{\alpha}(t)\big) + \mathcal{O}(\delta \,N \Vert X \Vert).
\end{eqnarray}
Apparently, if $\delta \, N\ll 1$ then, to a good approximation, the state $\omega_{\,t}$ is an incoherent superposition of eigenstates of the disjoint projections $Q_{\alpha}(t), \, \alpha = 1,\dots, N$\,. We then say that, at approximately time $t$, ``a projective (direct) measurement of $\mathcal{Q}$ takes place''.\\

\underline{\bf{Definition 5}}. \textit{(Resolution of $\mathcal{Q}$ in recording an event)}\\
Assuming that $\mathcal{X}_{t}$ is a countable set, then, for any $\delta \in (0,1)$, there exists a subset $\mathcal{X}_{t}^{(M)} \subseteq \mathcal{X}_t$ whose cardinality is given by a finite integer $M$ such that 
$$\omega_{t}\Big( \sum_{\xi \in \mathcal{X}_{t}^{(M)}} \pi_{t, \xi} \Big) \geq 1-\delta\,.$$
Then, for an arbitrary operator $X\in \mathcal{E}_{\geq t}$,
$$\omega_{t}(X)= \sum_{\xi \in \mathcal{X}_{t}^{(M)}} \omega_{t} \big(\pi_{t,\xi}\,X\, \pi_{t,\xi} \big) + \mathcal{O}(\delta\,\Vert X \Vert)\,.$$
The \textit{``resolution''} of $\lbrace Q_{\alpha}(t) \rbrace_{\alpha=0}^{N} \subset \mathcal{Q}$ in recording the event \mbox{$\lbrace \pi_{t,\xi}, \xi \in \mathcal{X}_{t} \rbrace$} starting to happen at time $t$ is defined by
\begin{equation}\label{resolution}
\hspace{1cm} \mathfrak{R}:= \frac{N}{M}\cdot(1-\delta) \,, \,\,\text{for  }\, 2\leq N \leq M\,,\,\,\, (\mathfrak{R}=0, \, \text{for   }\, N=1)\,. \hspace{0.6cm} \square
\end{equation}

It turns out that property (c), Eq. \eqref{dist}, above, implies that, given an orthogonal projection
 $Q_{\alpha}(t) \in \sigma_{t}^{\mathcal{Q}}(\mathcal{Q})$, there exists an orthogonal projection 
 \mbox{$P_{\alpha} \in \mathcal{Z}_{\omega_{\,t}}\big( \mathcal{E}_{\geq t} \big)$} such that
 \begin{equation} \label{proj}
 \Vert Q_{\alpha}(t) - P_{\alpha} \Vert < \mathcal{O}(\delta)\,.
 \end{equation}
 A proof of this simple lemma can be found in the appendix of \cite{FS-Vienna}.
 
 Since the projections $\pi_{t,\xi}, \xi \in \mathcal{X}_t$ generate the abelian algebra $\mathcal{Z}_{\omega_{\,t}}\big(\mathcal{E}_{\geq t}\big)$, we have that 
 \begin{equation}\label{range}
 \pi_{t, \xi}\cdot P = \pi_{t, \xi} , \, \text{ or }\, \pi_{t,\xi}\cdot P =0, \quad \forall \xi \in \mathcal{X}_t\,, 
 \end{equation}
 for any orthogonal projection $P\in \mathcal{Z}_{\omega_{\,t}}\big(\mathcal{E}_{\geq t}\big)$.
Equations \eqref{proj} and \eqref{range} then imply the \vspace{0.2cm}\\
\underline{\bf{Result}}.
For any $\alpha =1,\dots, N$, and for all $\xi \in \mathcal{X}_{t}$, 
$$\boxed{ \,\,\Vert \pi_{t, \xi} \, Q_{\alpha}(t) - \pi_{t, \xi} \Vert < \mathcal{O}(\delta)\,, \,\, \text{or  }\,\, \Vert \pi_{t, \xi} Q_{\alpha}(t) \Vert < \mathcal{O}(\delta)\,.}$$

Suppose that the physical quantity $\mathcal{Q}$ is generated by all functions of a single self-adjoint operator $\hat{Y}$. Then the best estimate for the value of $\hat{Y}$ right after time $t$ when the event 
$\lbrace \pi_{t,\xi} \vert \xi \in \mathcal{X}_{t} \rbrace$ has started to happen is an eigenvalue of $\hat{Y}$ corresponding to an eigenstate of the operator $Y(t)\equiv \sigma_{\,t}^{\mathcal{Q}}(\hat{Y})$ in the range of the projection $Q_{\alpha}(t)$. The state of $S$ right after time $t$ is then given by 
$$[\omega_{\,t}(\pi_{t, \xi_{\,\flat}})]^{-1} \omega_{\,t}\big( \pi_{t,\xi_{\,\flat}} (\cdot) \pi_{t,\xi_{\,\flat}}\big)\,,$$
for some $\xi_{\,\flat} \in \mathcal{X}_{t}$ for which 
\begin{equation} \label{recording}
\Vert \pi_{t,\xi_{\,\flat}} Q_{\alpha}(t) - \pi_{t, \xi_{\,\flat}} \Vert < \mathcal{O}(\delta)\,.
\end{equation}
Furthermore:\\
The higher the resolution, $\mathfrak{R}$, of $\mathcal{Q}$ in recording the event
$\lbrace \pi_{t,\xi}, \xi \in \mathcal{X}_{t}\rbrace$, the more precise the information provided by a measurement of $\mathcal{Q}$ is; if $N=M$ and 
$\delta$ is sufficiently small then every $\widehat{Q}_{\alpha}$ determines a unique point $\xi_{\flat} \in \mathcal{X}_t$ with the property that $\Vert Q_{\alpha}(t) - \pi_{t, \xi_{\flat}} \Vert < \mathcal{O}(\delta)$. (In the limit where $\delta \rightarrow 0$ the information on the event that starts to happen at time $t$ becomes totally accurate.)

\textit{Remarks:}\\ 
\textit{(1)} The main results of this paragraph are Eq. \eqref{superpos}, the {\bf{Result}} stated above, and Eq. \eqref{recording}.
\\
\textit{(2)} The concepts presented in paragraph V. and results closely related to the ones described above can be obtained without ever using the theory of conditional expectations. However, their use renders the presentation more elegant.
}
\end{enumerate}
This completes our review of the ``$ETH$-Approach to Quantum Mechanics'' in a non-relativistic setting. Some idealized models fitting into this framework are discussed elsewhere, \cite{Les-Diablerets}. A relativistic form of this approach will be presented in \cite{Fr}. The material in \cite{Fr} leads one to speculate that a logically coherent quantum theory of events, measurements and observations in \textit{realistic} autonomous isolated (open) systems -- not involving the intervention of ``observers'' -- can only be developed in the realm of \textit{local relativistic quantum theories} with \textit{massless} particles, and for even-dimensional space-times.

\section{Scattered remarks about indirect measurements, conclusions}
I start this section with a few comments on ``indirect measurements'' (see \cite{Kraus, Maassen-Kummerer} for important early results) and then sketch some conclusions.

Let $S$ be an isolated open system, as discussed in Sections 2 and 3. I assume that the system has been prepared in such a way that there is a specific physical quantity, $\mathcal{Q}$, characteristic of $S$ that repeatedly records events featured by $S$ (i.e., is ``measured projectively''), at times $t_1<t_2<\dots < t_n, n\in \mathbb{N}$, as discussed in paragraph V. of Section 4, Eqs. \eqref{superpos} and \eqref{recording}. Let us assume that the spectrum of $\mathcal{Q}$ is a finite set $\mathcal{Y}^{\mathcal{Q}} = \lbrace 0,1, \dots, k \rbrace$, so that 
$\mathcal{Q}$ is generated by a single self-adjoint operator, $\widehat{Y}$, with eigenvalues $0, 1,2,\dots,k$. Let
\begin{equation}\label{meas-prot}
\underline{\eta}^{(n)}:= \lbrace \eta_1, \eta_2, \dots, \eta_n \rbrace, \quad \eta_{j} \in \mathcal{Y}^{\mathcal{Q}}\,, \,\, j=1,2,\dots, n\,,
\end{equation}
be the sequence of values of $\widehat{Y}$ measured at times $t_1, t_2, \dots, t_n$, as explained in paragraph V. of Section 4. This means that the state of $S$ right after time $t_j$ is in an approximate eigenstate corresponding to the eigenvalue $\eta_j$ of the operator $Y(t_j)$ representing 
$\widehat{Y}$ at time $t_j$, for $j=1,2,\dots,n$, as expressed in Eq. \eqref{superpos}. The sequence $\underline{\eta}^{(n)}$ is called a ``measurement protocol'' of length $n$. As an example, $\widehat{Y}$ may describe the functioning of $k$ different detectors that click when a certain type of particle (e.g., a photon or an atom), called a \textit{``probe''}, belonging to $S$ impacts them, with the following meaning of its eigenvalues:
\begin{center}
$\quad \eta =0\, \leftrightarrow \text{   none of the detectors clicks  },\, \eta = \ell \, \leftrightarrow \text{   detector  } \ell \,\text{  has clicked}\,,$\\
$\ell = 1,\dots, k$.
\end{center}
Given a measurement protocol $\underline{\eta}^{(n)}$ of length $n$, we define the frequency (of occurrence) of the value 
$\eta \in \mathcal{Y}^{\mathcal{Q}}$ by
\begin{equation}\label{frequency}
f_{\eta}\big(\underline{\eta}^{(n)}\big):= \frac{1}{n} \Big(\sum_{j=1}^{n} \delta_{\eta\, \eta_j} \Big)\,.
\end{equation}
Note that
$$f_{\eta}\big(\underline{\eta}^{(n)}\big) \geq 0, \quad \,\text{and   }\,\, \sum_{\eta=1}^{k} f_{\eta}\big(\underline{\eta}^{(n)}\big) = 1\,.$$
Of particular interest is the asymptotics of $f_{\eta}\big(\underline{\eta}^{(n)}\big)$, as $n\rightarrow \infty$. Let us temporarily assume that, $\forall \eta = 0, 1,\dots, k$,
the limit of $f_{\eta}\big(\underline{\eta}^{(n)}\big)$, as $n\rightarrow \infty$, exists whenever a copy of $S$ prepared in a fixed initial state is subjected to very many repeated measurements of $\widehat{Y}$, with
\begin{equation}\label{asymptotics}
\underset{n \rightarrow \infty}{\text{lim}}\, f_{\eta}\big(\underline{\eta}^{(n)}\big) \in \lbrace p(\eta\vert \alpha)\rbrace_{\alpha = 1}^{N}\,, 
\end{equation}
for some $N<\infty$; (this is a \textit{``Law of Large Numbers''}, see \cite{BFFS}). In \eqref{asymptotics},
\begin{equation}\label{p}
p(\eta \vert \alpha) \geq 0 , \, \text{   and   }\, \sum_{\eta=1}^{k} p(\eta \vert \alpha) =1\,,
\end{equation}
for all $\alpha=1,\dots, N$, for some $N<\infty$. Apparently, the probability measures $p(\cdot \vert \alpha), $ $\alpha = 1,\dots, N,$ describe all possible limiting values the frequencies $f_{(\cdot)}(\underline{\eta}^{(n)})$ may converge to. We propose to interpret the parameter $\alpha$ as follows: $\alpha$ characterizes a \textit{time-independent} property of $S$, i.e., it is an eigenvalue of a self-adjoint operator, $A$, on $\mathcal{H}$ representing a physical quantity of $S$ that commutes with the operators $Y(t_j), j=1,2,\dots,$ and is a 
\textit{conservation law}, meaning that $A$ is time-independent (under the Heisenberg time evolution of operators on 
$\mathcal{H}$). Such an indirect measurement of $A$ is called a \textit{``non-demolition measurement''}. One expects that conservation laws are elements of 
$$\mathcal{E}_{\infty} := \bigwedge_{t\in \mathbb{R}} \mathcal{E}_{\geq t}\,,$$
where $\mathcal{E}_{\infty}$ is an algebra in the center of the algebra $\mathcal{E}$ defined in \eqref{allevents}, (``asymptotic abelianness'' in time). Under suitable hypotheses this expectation can actually be proven. \\
Thus, if the frequencies $f_{\eta}\big(\underline{\eta}^{(n)}\big)$ are seen to converge to the value $p(\eta \vert \alpha_{*})$, as $n \rightarrow \infty$, $\eta \in \mathcal{Y}^{\mathcal{Q}}$, for some $\alpha_{*} \in \text{spec}(A)$, and if the measures $p(\cdot\vert \alpha)$ separate points in the spectrum, spec$(A)$, of $A$, then we \textit{know} that, asymptotically, as $t \rightarrow \infty$, the value of the conservation law $A$ approaches $\alpha_{*}$. (The fact that the measures 
$p(\cdot \vert \alpha)$ may depend on $\alpha$ in a non-trivial way, at all, is a consequence of \textit{``entanglement''}; see \cite{Maassen-Kummerer, Bauer-Bernard, BFFS}.)\\
Evidently, one would like to prove \eqref{asymptotics} and to predict the probability of indirectly measuring a value $\alpha_{*}$ for $A$, assuming one knows the initial state of $S$. However, this can only be done if the events encoded by the values 
$\eta_1, \eta_2, \dots$, of the physical quantity $\hat{Y}$, which is measured at times $t_1,t_2, \dots$, are the \textit{only} events happening in $S$. For a limited class of systems (see \cite{Bauer-Bernard, BFFS}), one can prove that if this is the case then \eqref{asymptotics} holds, the state of $S$ approaches an eigenstate of $A$ corresponding to some eigenvalue $\alpha_{*}\in \text{spec}(A)$, as time $t\rightarrow \infty$, (\textit{``purification''}), and the probability of measuring the value $\alpha_{*}$ is given by \textit{Born's Rule} applied to the initial state of $S$ and the operator $A$, see \cite{BFFS}.

Usually, operators on $\mathcal{H}$ representing physical quantities of $S$ are \textit{not} time-independent. If the rate of change in time of a physical quantity, $A$, of $S$ that one attempts to measure \textit{indirectly}, as described above, is very \textit{small}, as compared to the rate of repeated projective measurements of the physical quantity $\hat{Y}$ used to determine the value of $A$,\footnote{One speaks of a ``weak measurement'' of $A$} then it turns out that, to good accuracy, the dynamics of the state of the system $S$ is described by a \textit{Markov jump process} on the set of eigenspaces of the operator 
$A$. The sample paths of this process describe {\bf{``quantum jumps''}} of (the state of) $S$ from one approximate eigenstate
 of $A$ to another one. This picture has been given a precise meaning in \cite{BFFS, BCFFS-2}, in the framework of some simple models.\\

\underline{\bf{Concluding Remarks}}:
\begin{enumerate}
\item[(1)]{The $ETH$-Approach to $QM$ sketched in this paper is a ``Quantum Mechanics without observers''. It introduces a precise notion of ``events'' into the quantum formalism; and it furnishes quantum theory with a clear ``ontology''.}
\item[(2)]{The $ETH$-Approach establishes a precise formalism to describe the \textit{stochastic time evolution of states} of isolated (open) systems featuring events. As I have tried to explain, while, for an \textit{isolated} system, the Heisenberg-picture time evolution of operators, in particular of physical quantities characteristic of such a system, determined by the unitary propagator of the system is perfectly adequate, the time evolution of its \textit{states} is described by a novel kind of \textit{stochastic branching process} with a ``non-commutative state space''. This is described in some detail in paragraph IV. of Section 3. The analysis presented there shows that it is simply \textit{not true} -- in any naive sense -- that the ``Heisenberg picture'' and the ``Schr\"{o}dinger picture'' are equivalent.}
\item[(3)]{It is explained in paragraph V. of Section 3 what a ``physical quantity'' characteristic of an isolated open system is, what it means to measure such a quantity ``projectively'', and how ``projective measurements'' of physical quantities can be used to record events. This also lays a basis for a precise \textit{theory of indirect measurements}.}
\item[(4)]{It is important to note that, in the $ETH$-Approach to $QM$, the expected value of a \textit{conservation law} represented by a self-adjoint operator $A$ in the actual state of an isolated open system featuring events is {\bf{not}} constant in time, (as it would be if states evolved according to the Schr\"{o}dinger equation).}
\item[(5)]{A \textit{``passive state''} of an isolated open system $S$ prepared at some time $t_0$ is a state $\omega$ for which 
$\mathcal{Z}_{\omega_{\,t}}(\mathcal{E}_{\geq t}) = \lbrace \mathbb{C}\, \mathds{1} \rbrace\,,$ for all times $t>t_0$. We expect that it often happens that states of $S$ approach ``passive states'' asymptotically, as $t\rightarrow \infty$, (with $\sigma(\mu_{\omega})=0$, see \eqref{spec-entropy}). Thermal equilibrium states are ``passive states''.
}
\item[(6)]{Clearly, the $ETH$-Approach to $QM$ is so general that, for the time being, it is very hard to use it to carry out explicit calculations for \textit{realistic} model systems and to show in which way its predictions \textit{differ} -- usually (hopefully) only ever so slightly -- from those made on the basis of, for example, the Copenhagen Interpretation of $QM$, or \textit{Bohmian Mechanics}. I emphasize, however, that \textit{differences in the predictions of the $ETH$-Approach and other versions of $QM$ -- however small they may be -- really exist!}}
\item[(7)]{After completion of this work \textit{Bernard Kay} has pointed out to me that in two of his papers -- see \cite{Kay} -- ideas somewhat related to some of the ideas proposed in the present paper have been described. I thank Bernard for valuable discussions.}

\end{enumerate}


\begin{thebibliography}{References}

\bibitem{Griffiths}R. B. Griffiths, \textit{Consistent Histories and the Interpretation of Quantum Mechanics}, J. Stat. Phys. {\bf{36}} (1984), 219-272;\\
M. Gell-Mann and J. B. Hartle, \textit{Classical Equations for Quantum Systems}, Phys. Rev. D{\bf{47}} (1993), 3345-3382 

\bibitem{Durr-Teufel} D. D\"{u}rr and S. Teufel, \textit{Bohmian Mechanics -- The Physics and Mathematics of Quantum Theory}, Springer-Verlag, Berlin-Heidelberg-New York 2009

\bibitem{FS-Vienna}J. Fr\"{o}hlich and B. Schubnel, \textit{Do We Understand Quantum Mechanics -- Finally?}, arXiv:1203.3678, in: Proceedings of conference in memory of Erwin Schr\"{o}dinger, Vienna, January 2011, publ. in 2012

\bibitem{Abou-Salem-F} W.K. Abou Salem and J. Fr\"{o}hlich, \textit{Status of the Fundamental Laws of Thermodynamics}, J. Stat. Phys. {\bf{126}} (2007), 1045–1068

\bibitem{DeR-Fr} W. De Roeck and J. Fr\"{o}hlich, \textit{Diffusion of a Massive Quantum Particle Coupled to a Quasi-Free Thermal Medium}, Commun. Math. Phys. {\bf{303}} (2011), 613-707

\bibitem{Gang-F} J. Fr\"{o}hlich, Zhou Gang and A. Soffer, \textit{Friction in a Model of Hamiltonian Dynamics}, Commun. Math. Phys. {\bf{315}} (2012), 401-444;\\
J. Fr\"{o}hlich and Zhou Gang, \textit{Emission of Cherenkov Radiation as a Mechanism for Hamiltonian Friction}, Adv. Math. {\bf{264}} (2014), 183-235

\bibitem{B-DeR-F} R. Bauerschmidt, W. De Roeck and J. Fr\"{o}hlich, \textit{Fluctuations in a Kinetic Transport Model for Quantum Friction}, arXiv:1403.5790, J. Physics A: Math. Theor. {\bf{47}} (2014), 275 003

\bibitem{FS-Prob-Theory} J. Fr\"{o}hlich and B. Schubnel, \textit{Quantum Probability Theory and the Foundations of Quantum Mechanics}, arXiv:1310.1484, in: \textit{The Message of Quantum Science -- Attempts Towards a Synthesis}, Ph. Blanchard and J. Fr\"{o}hlich (eds.), Springer-Verlag, Berlin-Heidelberg-New York 2015

\bibitem{BFS-forks} Ph. Blanchard, J. Fr\"{o}hlich and B. Schubnel, \textit{A 'Garden of Forking Paths' -- the Quantum Mechanics of Histories of Events}, Nucl. Phys. B{{\bf912}} (2016), 463-484

\bibitem{Schubnel-thesis} B. Schubnel, \textit{Mathematical Results on the Foundations of Quantum Mechanics}, PhD thesis 2014, available at https://doi.org/10.3929/ethz-a-010428944

\bibitem{Les-Diablerets} J. Fr\"{o}hlich, \textit{ 'ETH' in Quantum Mechanics}, Notes of Lectures on the Foundations of Quantum Mechanics, Les Diablerets, January 9 - January 14, 2017

\bibitem{FFS} J. Faupin, J. Fr\"{o}hlich and B. Schubnel, \textit{On the Probabilistic Nature of Quantum Mechanics and the Notion of 'Closed' Systems}, Ann. Henri Poincar\'e {\bf{17}} (2016), 689-731

\bibitem{FS-state-prep} J. Fr\"{o}hlich and B. Schubnel, \textit{The Preparation of States in Quantum Mechanics}, J. Math. Phys. {\bf{57}} (2016), 042 101

\bibitem{Haag} R. Haag, \textit{Fundamental Irreversibility and the Concept of Events}, Commun. Math. Phys. {\bf{132}} (1990), 245-251;\\
R. Haag, \textit{Events, Histories, Irreversibility}, in: \textit{Quantum Control and Measurement}, Proc. ISQM, ARL Hitachi, H. Ezawa and Y. Murayama (eds.), North Holland, Amsterdam 1993;\\
Ph. Blanchard and A. Jadczyk, \textit{Event-Enhanced Quantum Theory and Piecewise Deterministic Dynamics}, Annalen der Physik {\bf{4}} (1995), 583-599

\bibitem{Buchholz} D. Buchholz and J. E. Roberts, \textit{New Light on Infrared Problems: Sectors, Statistics, Symmetries and Spectrum}, Commun. Math. Phys. {\bf{330}} (2014), 935-972;\\
D. Buchholz, \textit{Collision Theory for Massless Bosons}, Commun. Math. Phys. {\bf{52}} (1977), 147-173

\bibitem{Fr} J. Fr\"{o}hlich, \textit{Quantum Theory and Causality}, Talks at the University of Leipzig (2018), TU-Stuttgart (2019), IHES (2019) and at Vietri sul Mare (Italy) (2019); paper in preparation

\bibitem{Raimond-Haroche} C. Guerlin, J. Bernu, S. Del\'eglise, C. Sayrin, S. Gleyzes, S. Kuhr, M. Brune, J. M. Raimond and S. Haroche, \textit{Progressive Field-State Collapse and Quantum Non-Demolition Photon Counting}, Nature {\bf{448}} (7156) (2007), 889-893;\\
S. Haroche, \textit{Controlling Photons in a Box and Exploring the Quantum to Classical Boundary}, Nobel Lecture, December 8, 2012, in: \textit{The Nobel Prizes}

\bibitem{Bauer-Bernard} M. Bauer and D. Bernard, \textit{Convergence of Repeated Quantum Non-Demolition Measurements and Wave-Function Collapse}, Phys. Rev. A{\bf{84}} (2011), 044 103:1-4

\bibitem{Maassen-Kummerer} H. Maassen and B. K\"{u}mmerer, \textit{Purification of Quantum Trajectories}, Springer-Verlag, Lecture Notes - Monograph Series {\bf{48}} (2006), 252-261 

\bibitem{BFFS} M. Ballesteros, M. Fraas, J. Fr\"{o}hlich and B. Schubnel, \textit{Indirect Retrieval of Information and the Emergence of Facts in Quantum Mechanics}, arXiv:1506.01213, J. Stat. Phys. {\bf{162}} (2016), 924-958 

\bibitem{BCFFS-1} M. Ballesteros, N. Crawford, M. Fraas, J. Fr\"{o}hlich and B. Schubnel, \textit{Non-Demolition Measurements of Observables with General Spectra}, arXiv:1706.09584, Proceedings publ. in 2018

\bibitem{BCFFS-2} M. Ballesteros, N. Crawford, M. Fraas, J. Fr\"{o}hlich and B. Schubnel, \textit{Perturbation Theory for Weak Measurements in Quantum Mechanics, I - Systems with Finite-Dimensional States Space}, arXiv1700.03149, Ann. Henri Poincar\'e {\bf{20}} (2019), 299-335 

\bibitem{Bell-book} J. S. Bell, \textit{Speakable and Unspeakable in Quantum Mechanics}, Cambridge University Press, Cambridge (UK) 1987.\\
See also:\\
J. A. Wheeler and W. H. Zurek, \textit{Quantum Theory and Measurement}, Princeton University Press, Princeton NJ, 1983;\\
K. Hepp, \textit{Quantum Theory of Measurement and Macroscopic Observables}, Helv. Phys. Acta {\bf{45}} (1972), 237-248;\\
H. Primas, \textit{Asymptotically Disjoint Quantum States}, in: \textit{Decoherence: Theoretical, experimental and Conceptual Problems}, pp 161-178, Ph. Blanchard, D. Giulini, E. Joos, C. Kiefer and I.-O. Stamatescu (eds.), Springer-Verlag, Berlin 2000

\bibitem{Renner} D. Frauchiger and R. Renner, \textit{Quantum Theory Cannot Consistently Describe the Use of Itself}, Nature Communications {\bf{9}} (2018), \# 3711 

\bibitem{Schwinger} G. L\"{u}ders, \textit{\"{U}ber die Zustands\"{a}nderung durch den Messprozess}, Annalen der Physik (Leipzig) {\bf{443}} (1950), 322-328;\\
J. Schwinger, \textit{The Algebra of Microscopic Measurement}, Proc. Natl. Acad. Sci. (USA) {\bf{45}} (1959), 1542-1553;\\
E. P. Wigner, \textit{The Collected Works of Eugene Paul Wigner, Part A: The Scientific Papers}, B. R. Judd, G. W. Mackey (eds.), Springer-Verlag, New York 1993

\bibitem{Takesaki} M. Takesaki, \textit{Conditional Expectations in von Neumann Algebras}, J. Funct. Anal. {\bf{9}} (1972), 306-321;\\
F. Combes, \textit{Poids et Esp\'erances Conditionnelles dans les Alg\`ebres de von Neumann}, Bull. Soc. Math. France {\bf{99}} (1971), 73-112

\bibitem{Kraus} K. Kraus, \textit{States, Effects and Operations}, Springer-Verlag, Berlin-Heidelberg-New York 1983

 \bibitem{Kay} Bernard S. Kay and Varqa Abyaneh, \textit{Expectation values, experimental predictions, events and entropy in quantum gravitationally decohered quantum mechanics}, arXiv:0710.0992 (v1), unpublished;\\
Bernard S. Kay, \textit{The Matter-Gravity Entanglement Hypothesis}, Found Phys {\bf{48}} (2018), 542-557


\end{thebibliography}
\end{document}